\documentclass[twocolumn,showpacs]{revtex4}
\usepackage{psfig}
\begin{document}

 \title{\hfill {\footnotesize FZJ--IKP(TH)--2004--20, HISKP-TH-04-24}\\
  Near threshold enhancement of the  
 {\boldmath $p{\bar p}$} mass spectrum in $J/\Psi$ decay}

%\title{Near threshold enhancement of the  
%{\boldmath $p{\bar p}$} mass spectrum in $J/\Psi$ decay}

\author{A. Sibirtsev$^{1}$, J. Haidenbauer$^2$, 
S. Krewald$^2$, Ulf-G. Mei{\ss}ner$^{1,2}$, and A.W. Thomas$^3$}

\affiliation{
$^1$Helmholtz-Institut f\"ur Strahlen- und Kernphysik (Theorie), 
Universit\"at Bonn, Nu\ss allee 14-16, D-53115 Bonn, Germany \\
$^2$Institut f\"ur Kernphysik (Theorie), Forschungszentrum J\"ulich,
D-52425 J\"ulich, Germany,\\
$^3$Jefferson Lab, 12000 Jefferson Ave., Newport News, VA 23606, USA
}

\begin{abstract}
We investigate the nature of the near-threshold enhancement in the 
$p{\bar p}$ invariant mass spectrum of the reaction
$J/\Psi{\to}\gamma p{\bar p}$ reported recently by the BES Collaboration. 
Using the J\"ulich $N\bar N$ model we show that the mass dependence of 
the $p{\bar p}$ spectrum close to the threshold can be reproduced by the 
$S$-wave $p{\bar p}$ final state interaction in the isospin $I{=}1$ state 
within the Watson-Migdal approach. 
However,
because of our poor knowledge of the $N\bar N$ interaction 
near threshold and of the $J/\Psi{\to}\gamma p{\bar p}$ 
reaction mechanism
and in view of the controversal situation in the decay 
$J/\Psi{\to}\pi^0 p{\bar p}$, where no obvious signs of a 
$p\bar p$ final state interaction are seen, explanations other than
final state interactions cannot be ruled out at the present stage.   
\end{abstract}

\maketitle

\section{Introduction}
Recently the BES Collaboration~\cite{Bai} reported a 
near-threshold enhancement in the proton-antiproton ($p{\bar p}$) invariant 
mass spectrum, observed in the $J/\Psi{\to}\gamma p{\bar p}$ decay. 
Signs for a low mass $p{\bar p}$ enhancement had
been already seen earlier by the Belle Collaboration in their
study of the $B^+{\to}K^+p{\bar p}$ decay~\cite{Abe1}
as well as in the reaction ${\bar B}^0{\to}D^0p{\bar p}$~\cite{Abe2}.
But because of the large statistical uncertainties of the Belle data it
was difficult to draw concrete, quantitative conclusions about 
the extent of the near threshold $p{\bar p}$ enhancement.
The new data by Bai et al.~\cite{Bai}, however, are of rather 
high statistical accuracy and therefore provide very precise 
information about the magnitude and the energy dependence of the $p{\bar p}$ 
mass spectrum very close to threshold.  
 
The BES Collaboration \cite{Bai} fitted their $p{\bar p}$ invariant mass
spectrum below 1.95~GeV by a Breit-Wigner resonance function. 
Assuming that the $p{\bar p}$ system is in an $S$-wave  
resulted in a resonance mass of $M{=}1859^{+3+5}_{-10-25}$~MeV
and a total width of $\Gamma{<}$30~MeV. A comparable fit to the data
could be achieved with a $P$-wave Breit-Wigner function 
with $M{=}1876{\pm}0.9$ MeV and $\Gamma{=}4.6{\pm}1.8$ MeV.

The proximity of these resonance masses to the $p\bar p$ 
reaction threshold (which is at 1876.54 MeV) nourished speculations 
that the observed strong enhancement could be a signal of an $N\bar N$
bound state. While theoretical considerations of such $N\bar N$
bound states (or of baryonia, in general) abound in the 
literature~\cite{Myhrer,Shapiro,Jaffe,Dover,Dalkarov1,Richard} 
there is so far hardly any undisputed experimental information
on the existence of such states 
\cite{Bridges,Gray,Reidlberger}. Thus, the supposition
that one has found here independent and possibly even more convincing
evidence in support of $N\bar N$ bound states is certainly appealing. 

An alternative explanation put forward by the BES collaboration 
invokes similarities of the observed enhancement in the $p\bar p$ mass 
spectrum near threshold with the strong energy dependence of
the electromagnetic form factor of the proton around $\sqrt{s}{\approx}2m_p$, 
in the timelike region, as determined in the reaction 
$p\bar p{\to}e^+e^-$ \cite{Bardin}. In the latter case it was argued that
the sharp structure seen in the experiment could be caused by a
narrow, near-threshold vector-meson ($J^{PC}$=$1^{--}$) 
resonance \cite{Antonelli1} with $M{=}1870{\pm}10$ MeV and 
$\Gamma{=}10{\pm}5$ MeV. (See also Refs.~\cite{Krewald,Hammer} for
a pertinent discussion.) 
One should keep in mind, however, that the quantum numbers of
the $J/\Psi$ particle ($J^{PC}$=$1^{--}$) would restrict such
a resonance to occur in a pseudoscalar ($0^{--}$) or scalar 
($0^{+-}$) state -- should it be indeed responsible for the enhancement 
of the near-threshold
$p\bar p$ mass spectrum in the decay $J/\Psi{\to}\gamma p{\bar p}$.

An entirely different and much more conventional
interpretation of the observed enhancement was suggested in
several recent works~\cite{Kerbikov2,Zou,Bugg}. These 
authors argue that the enhancement is primarily due to the
final state interaction (FSI) between the produced proton and antiproton.
Specifically, it was shown within the scattering length approximation 
\cite{Kerbikov2}
that a calculation with a complex $S$-wave scattering length extracted 
from an effective-range analysis of $p{\bar p}$ scattering 
data can reproduce the shape of the $p{\bar p}$ mass distribution
close to the threshold.

In the present paper we analyse the near threshold enhancement in 
the $p{\bar p}$ invariant mass spectrum reported by the BES 
Collaboration 
utilizing a realistic model of the $N\bar N$ interaction \cite{Hippchen}. 
The elastic part of this model is the $G$-parity transform of the Bonn 
meson-exchange $NN$ potential, supplemented by a phenomenological complex 
potential to account for $N\bar N$ annihilation. 

As just mentioned, the investigations of Kerbikov et al. \cite{Kerbikov2}
as well as those of Bugg \cite{Bugg} rely on the 
scattering length approximation. But it remains unclear over which
energy range this approximation can provide a reliable representation
of the energy dependence of the $p\bar p$ amplitude. The enhancement
seen in the BES data extends up to invariant masses of $M(p\bar p) 
{\approx}$2 GeV, which corresponds to center-of-mass energies of around 
120 MeV in the $p\bar p$ system. It is obvious that the simple
scattering length approximation cannot be valid over such a large
energy region. Using the $p\bar p$ amplitude of our interaction
model (which describes the available $N\bar N$ scattering data up to
center-of-mass energies of 150 MeV~\cite{Hippchen}) 
we can examine to what extent the 
scattering length approximation can indeed reproduce the energy dependence 
of the $p{\bar p}$ scattering amplitude. Moreover, we can compare
the energy dependence induced by the full $p\bar p$ amplitude 
with the BES data over the whole energy range where the enhancement
was observed.  

A microscopic $N\bar N$ model has a further advantage. It yields
predictions for all possible spin- and isospin channels. Thus, 
we can investigate whether the energy dependence of the $^1S_0$ and 
$^3P_0$ $N\bar N$ scattering amplitudes in the $I{=}0$ and $I{=}1$
isospin channels is compatible with the BES data. 
(We use the nomenclature $^{(2I+1)(2S+1)}L_J$ but omit the isospin
index when referring to both isospin channels or to isospin averaged
results.) 
The analysis of Kerbikov et al. utilizes scattering lengths obtained
from a spin-averaged effective-range fit to the low-energy $N\bar N$
data. Those values might be strongly influenced or even dominated by
the $^3S_1$ $N\bar N$ partial wave, a channel which cannot contribute to 
the FSI in the $J/\Psi{\to}\gamma p{\bar p}$ decay because of 
charge-conjugation invariance. 

Let us finally mention that the work of Zou and Chiang \cite{Zou} does not
employ the scattering length approximation but uses the $K$-matrix 
formalism. However, their interaction consists only of one-pion exchange
and is therefore not  realistic because it does not take into account 
the most striking feature of low-energy $N\bar N$ scattering, namely 
annihilation. 
Recall that near the $p\bar p$ threshold the annihilation cross section
is twice as large as the elastic $p\bar p$ cross section \cite{Hippchen}. 

The paper is structured in the following way: In Sect.~II we briefly
review the $N\bar N$ potential model that is used in the present
analysis and examine the reliability of the 
scattering length approximation on the $N\bar N$ amplitude for the
partial waves relevant to the analysis of the BES data.
The possible influence of the Coulomb interaction is discussed as well. 
In Sect. III we provide details of our calculation of the 
near-threshold $p\bar p$ mass spectrum for the reaction 
and we compare our results with the measurement of the BES collaboration. 
Sect.~IV is devoted to a discussion of possible signals of 
$N\bar N$ bound states or (sub $p \bar p$ threshold) meson
resonances in the $p\bar p$ mass spectrum.
The paper ends with concluding remarks. 

\section{Proton-antiproton scattering at low energies}

\begin{figure}[t]
\vspace*{0mm}
\centerline{\hspace*{-3mm}\psfig{file=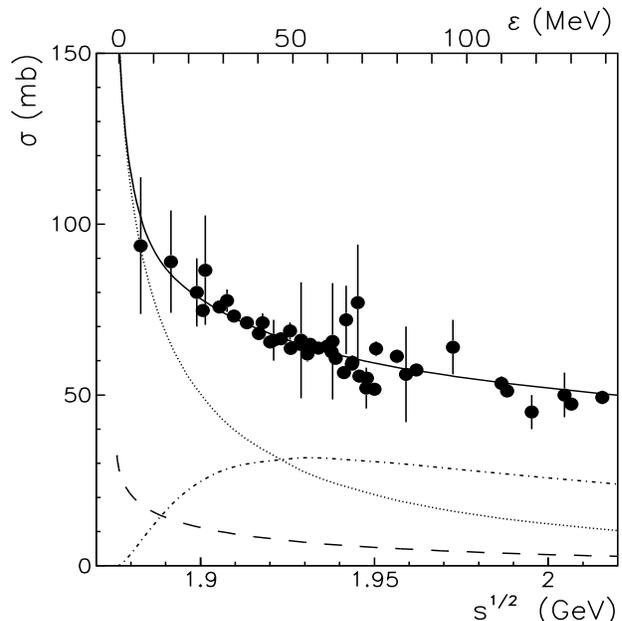,width=9.5cm,height=8.8cm}}
\vspace*{-7mm}
\caption{The $p{\bar p}$ elastic cross section as a
function of the invariant collision energy. The lines show the 
result of the J\"ulich model A(OBE)~\cite{Hippchen}. The dashed line
is the contribution from the $^1S_0$ partial wave alone, dotted - from
all $s$ waves, dashed-dotted from $p$ waves and solid lines is
full calculations. The circles are experimental data taken from 
Ref.~\cite{PDG}.}
\label{pbar7}
\end{figure}

In the present investigation we use one of the $N\bar N$ models
developed by the J\"ulich group \cite{Hippchen0,Hippchen,Mull}.
Specifically, we use the Bonn OBE (one-boson-exchange) model 
introduced in Ref.~\cite{Hippchen0}
whose results are also discussed in Ref.~\cite{Hippchen}. In the latter
work the model is called A(OBE) and we also adopt this 
name  in the present 
paper. The model A(BOX) is based on the OBE version of the full
Bonn potential derived in time-ordered perturbation theory, i.e. 
OBEPT in Ref.~\cite{MHE}. The G-parity transform of this interaction
model constitutes the elastic part of the $N\bar N$ potential. 
A phenomenological \hbox{spin-}, isospin- and energy-independent 
complex potential of Gaussian form is added to account for the 
$N\bar N$ annihilation. With only three free parameters (the range
and the strength of the real and imaginary parts of the annihilation
potential) a good overall description of the low- and intermediate 
energy $N\bar N$ data was achieved. Results for the total and the
integrated elastic and charge-exchange cross sections as well as
angular dependent observables can be found in Refs.~\cite{Hippchen0,Hippchen}. 
Here in Fig.~\ref{pbar7} we show only the elastic $p\bar p$ cross section.

We also utilized the most complete $N\bar N$ model of the J\"ulich 
Group, model D published in Ref. \cite{Mull}. The elastic part of this 
interaction model is derived from the G-parity transform of the full Bonn 
$NN$ potential and the annihilation is described in part microscopically by 
$N\bar N{\to}$ 2 meson decay channels - see Ref. \cite{Mull} for details.
But since the results turned out to be qualitatively rather similar to
the ones obtained with A(OBE) we refrain from showing them here. 

The differential cross sections for $p{\bar p}$ 
scattering~\cite{Bruckner1,Bruckner2,Linssen} already 
indicate a substantial 
contribution from higher partial waves  at around 10 MeV above 
the $p{\bar p}$ threshold. This feature is also reflected in the
predictions of the $N\bar N$ model. The dashed line in Fig. \ref{pbar7} shows 
the contribution from the $^1S_0$ partial wave. The dotted and dash-dotted 
lines illustrate the total contributions from $s$ and $p$ waves, 
respectively.

Since the scattering length approximation to the $N\bar N$ amplitude
was used in two of the analyses of the BES data \cite{Bugg,Kerbikov2}
we want to investigate
the validity of this approximation. In the following we ignore the
proton-neutron mass difference and also the Coulomb interaction in the
$p\bar p$ system, in order to simplify the discussion. But we will come
back to these issues later. 

In the scattering length approximation the $S$-wave 
$p{\bar p}$ scattering amplitude $T$ is given by 
\begin{eqnarray}
T=\frac{a}{1+iaq_p},
\label{scat1}
\end{eqnarray}
where the scattering length $a$ is a complex number because of inelastic
channels (annihilation into multimeson states) that are open already at 
the $p{\bar p}$ threshold.
The proton momentum in the c.m. system is $q_p=\sqrt{s-4m_p^2}/2$, 
where $m_p$ is the proton mass. 

\begin{figure}[t]
\vspace*{-1mm}
\centerline{\hspace*{2mm}\psfig{file=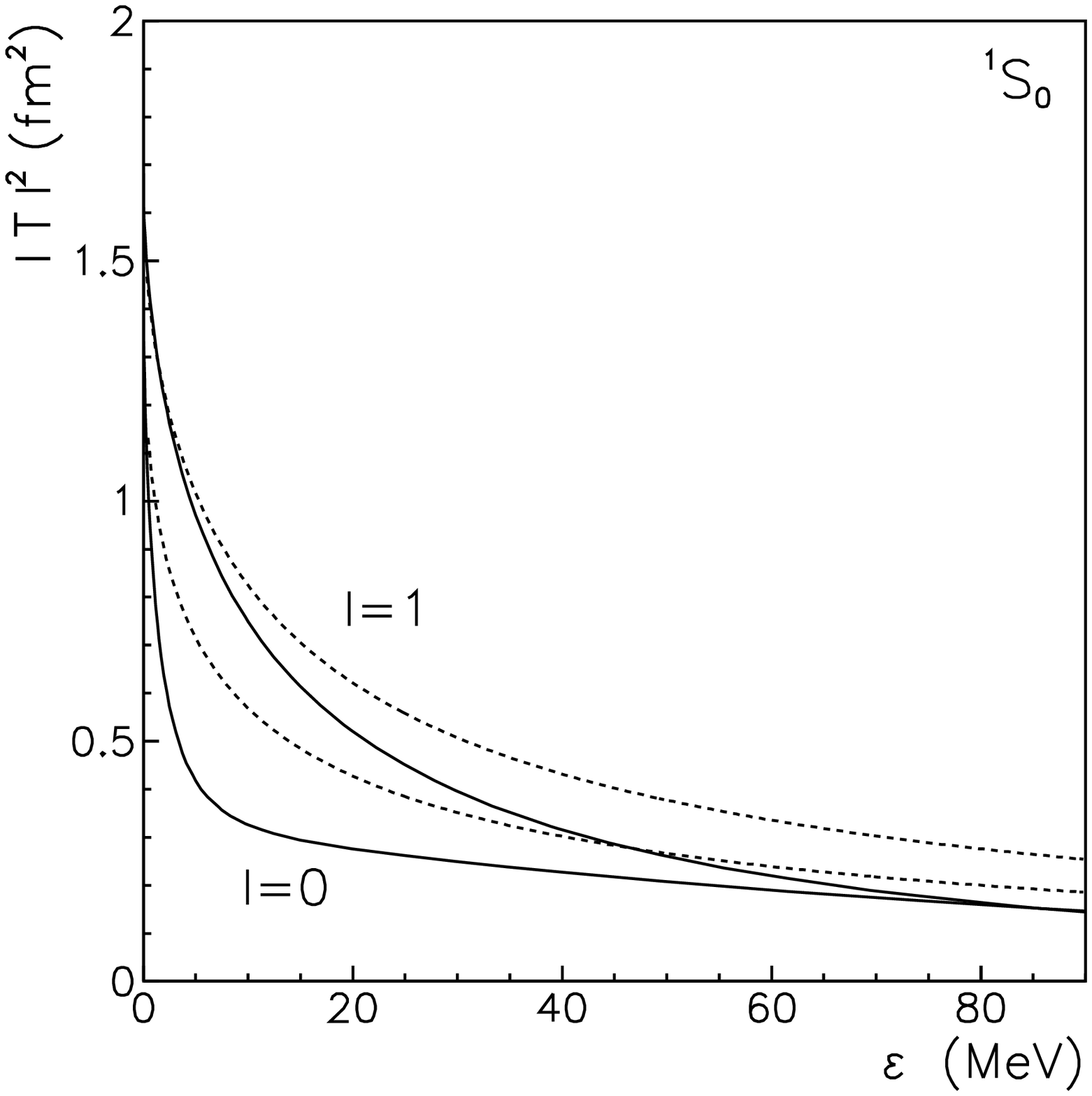,width=9.1cm,height=6.1cm}}
\vspace*{-5mm}
\centerline{\hspace*{2mm}\psfig{file=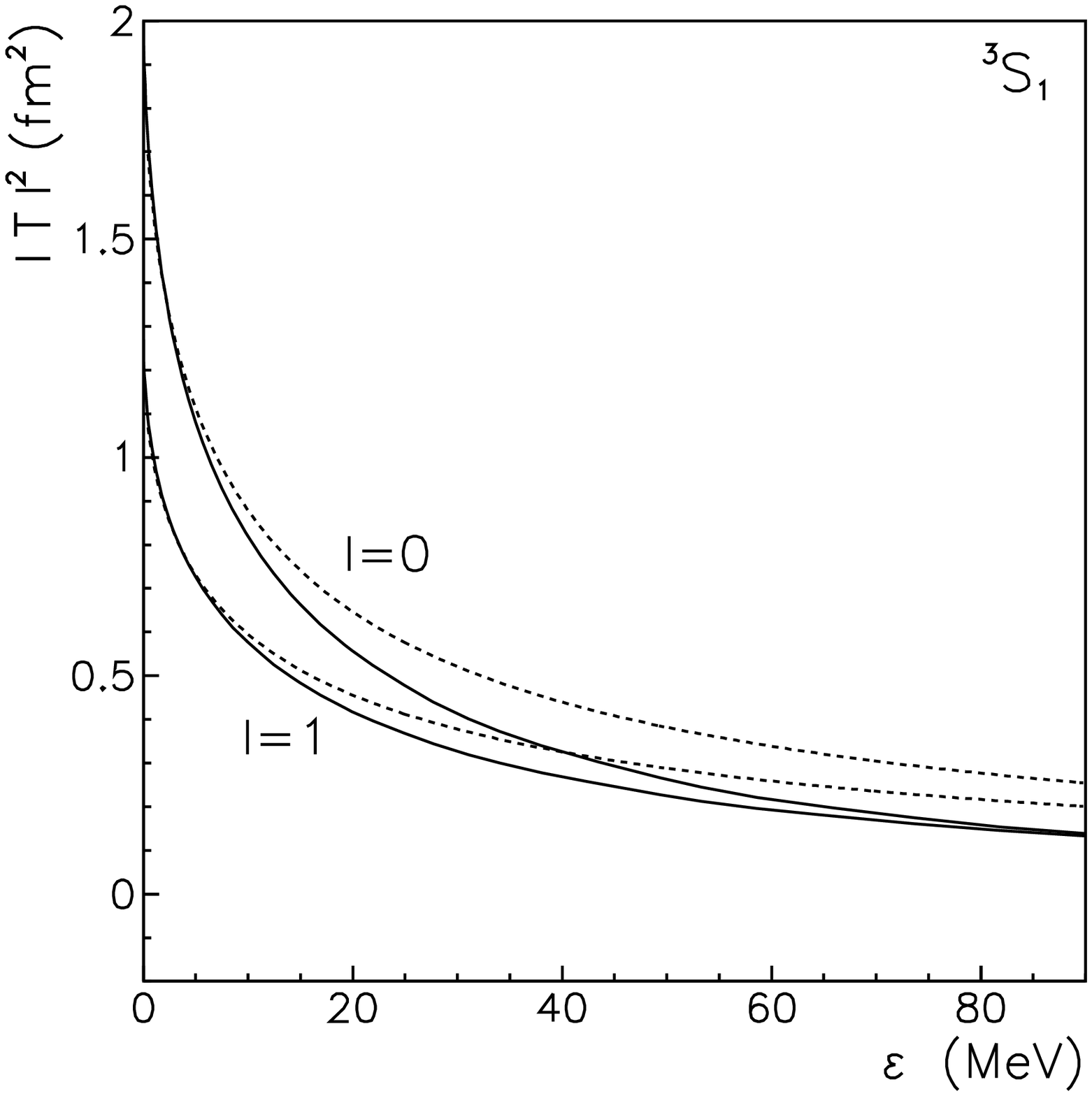,width=9.1cm,height=6.1cm}}
\vspace*{-5mm}
\caption{The $p{\bar p}$ scattering amplitudes for the $^1S_0$ and
$^3S_1$ partial waves as a function of the center-of-mass energy. 
The solid lines show the results of the
J\"ulich model while the dashed lines indicate the scattering length 
approximation given by Eq.(\ref{scat1}).}
\label{pbar6}
\end{figure}

Results for $|T|^2$ for the $^1S_0$ partial wave are presented in the upper
part of Fig.~\ref{pbar6}. The solid lines are the result for the full 
amplitude while the dashed lines are 
based on the scattering length approximation given by Eq.(\ref{scat1}). 
Note that the scattering lengths predicted by the 
$N\bar N$ model, which we use for the $^1S_0$ partial wave 
are $a_0{=}(-0.18{-}i1.18)$~fm 
and $a_1{=}(1.13{-}i0.61)$~fm for the isospin $I{=}0$ and $I{=}1$ 
channels, respectively. It is evident that the 
scattering length approximation does not reproduce the energy dependence
of the scattering amplitude that well. For the $I{=}1$ channel the difference
at an excess energy of 50 MeV amounts as much as 50 \%. 
The difference is even more pronounced for the $I{=}0$ channel, where 
we already observe large deviations from the full result at rather 
low energies. This strong failure of the scattering length approximation 
is due to the much smaller scattering length predicted by our model 
for the $I{=}0$ partial wave. 

Results for the $^3S_1$ partial wave are shown in the lower part of 
Fig.~\ref{pbar6}.
Here the scattering lengths predicted by the $N\bar N$ model are
$a_0{=}(1.16{-}i0.82)$~fm and $a_1{=}(0.75{-}i0.84)$~fm for the $I{=}0$ 
and $I{=}1$ channels, respectively. As pointed out above, this
partial wave cannot contribute to the reaction $J/\Psi{\to}\gamma p{\bar p}$.
But it is still interesting to see how well the scattering length 
approximation works in this case. Obviously we also see similar
shortcomings here. Thus, it is clear from Fig. \ref{pbar6} that
the use of the scattering length approximation allows only a very
rough qualitative estimate of the effects of  FSI but it 
is definitely not reliable for a more quantitative analysis of 
the BES data. 

\begin{figure}[t]
\vspace*{-1mm}
\centerline{\hspace*{-5mm}\psfig{file=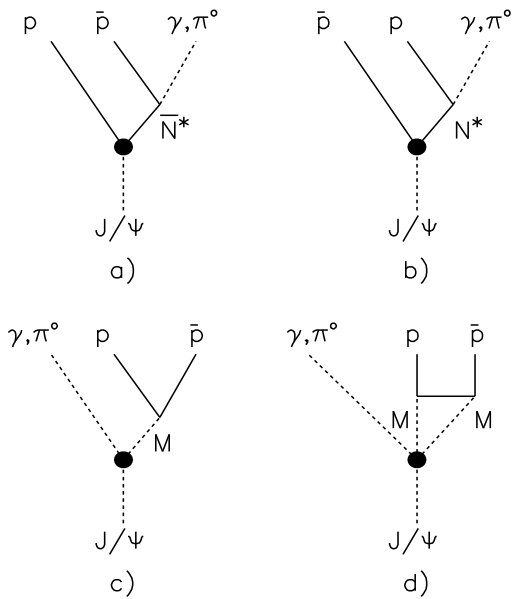,width=9.4cm,height=9cm}}
\vspace*{-7mm}
\caption{Some possible reaction mechanisms for the 
decays $J/\Psi{\to}\gamma p{\bar p}$ and 
$J/\Psi{\to}\pi^0 p{\bar p}$. $M$ indicates intermediate mesonic states
and $N$ ($\bar N$) intermediate nucleon (antinucleon) resonances. 
}
\label{diag}
\end{figure}

Nevertheless, one has to realize that the main uncertainty in estimating
$p\bar p$ FSI effects does not come from the scattering length approximation
but from our poor knowledge of the $p\bar p$ $^1S_0$ amplitudes near
threshold and of the $J/\Psi{\to}\gamma p{\bar p}$ reaction mechanism.
For example,
the scattering lengths employed by Kerbikov et al. are spin-averaged
values. Since the $^3S_1$ partial wave contributes with a weighting factor 3
to the $p\bar p$ cross sections (and there are no spin-dependent observables
at low energies that would allow one to disentangle 
the spin-dependence) it is obvious
that their value should correspond predominantly to the $^3S_1$ amplitude.
Thus, it is questionable whether it should be used for analyzing the BES
data at all because the contribution of the $^3S_1$ partial wave to
the decay $J/\Psi{\to}\gamma p{\bar p}$ is forbidden by charge-conjugation
invariance.  
But even the availability of genuine $^1S_0$ amplitudes (which
are necessarily model dependent) does not solve the problem. The reaction
$J/\Psi{\to}\gamma p{\bar p}$ can have any isospin combination in the final
$p\bar p$ state. A dominant reaction mechanism involving an intermediate
isoscalar meson resonance ($\eta$, $\eta_c$,~...), cf. Fig.~\ref{diag}c, 
would yield a pure $I{=}0$ final $p\bar p$ state, whereas
an intermediate isovector meson resonance ($\pi$, $\pi(1800)$,~...)
leads to pure $I{=}1$.  On the other hand, mechanisms like the ones 
depicted in Fig.~\ref{diag}a,b generate the usual
equal weighting of the two isospin amplitudes.

In this context let us mention that the spin-averaged $p\bar p$ 
scattering length predicted by the model A(BOX), $a_S{=} 0.84{-}i0.85$~fm,
is in rough agreement with the value extracted from the level shifts of
antiproton-proton atoms, cf. Table 8 in Ref.~\cite{Gotta}.
On the other hand, this averaged value differs significantly from 
the scattering
length in the $^{11}S_0$ partial wave 
-- which is one of the
possible final states in the decay $J/\Psi{\to}\gamma p{\bar p}$.
At first glance one would tend to believe that the unusually small
real part of $a_0$ is basically accidental and reflects the fact that
the $^1S_0$ state enters with smaller statistical weight into the
$p\bar p$ cross section and, therefore, is much less constrained by
the $p\bar p$ data. However, exploratory calculations with the other
$N\bar N$ models of the J\"ulich group revealed that the real part of
the $^{11}S_0$ scattering length is always small (and often of a different 
sign from the other $S$ waves). For example, the
more involved $N\bar N$ model D published in Ref. \cite{Mull},
yields $a_0{=}-0.25{-}i1.01$~fm. 
Moreover,  the $N\bar N$ models of Dover-Richard also
and Kohno-Weise predict rather small values for the real part of
the $^{11}S_0$ scattering length, as is evident from the results
presented in Ref.~\cite{Carbonell}. 
The Paris $N\bar N$ potential, on the other hand,
predicts the real parts of the $^{1}S_0$ scattering lengths to 
be of similar magnitude - cf. Table IV in Ref.~\cite{Paris}. But 
Fig.~11 of that paper suggests that the central potential in the
$I=0$, $S=0$ channel has been drastically modified inside
1 fm as compared to what follows from the
$G$-parity transformation of the Paris $NN$ potential. In the
J\"ulich $N{\bar N}$ models no such changes are introduced when 
deriving the elastic part of the $N{\bar N}$ interaction via
$G$-parity! 

A closer examination of our $N\bar N$ models disclosed that here the smallness
of $Re \ a_0$ is definitely caused by the one-pion exchange contribution,
i.e. by long-range physics. Switching off this contribution always led
to a large scattering length similar to the one in the $I{=}1$ channel,
whereas modifications of the short-range contributions to the 
elastic part of the $N\bar N$ potential had hardly any qualitative
influence on the scattering lengths. Thus, it would be interesting to 
obtain experimental constraints on the $^{1}S_0$ scattering lengths.
Corresponding measurements could be performed at the future GSI
facility FLAIR \cite{FLAIR}, where it is possible to 
have a polarized antiproton
beam \cite{PAX}, as required for inferring the spin-triplet
and spin-singlet amplitudes from the data. 

\section{$J/\Psi$ decay rate and FSI effects}

The $J/\Psi{\to}\gamma p{\bar p}$ decay rate is given 
as~\cite{Byckling}
\begin{eqnarray}
d\Gamma = \frac{|A|^2}{2^9 \pi^5 m_{J/\Psi}^2}\,
\lambda^{1/2}(m_{J/\Psi}^2,M^2,m_\gamma^2) \nonumber \\
\times\lambda^{1/2}(M^2,m_p^2,m_p^2)\, dM d\Omega_p\,  d\Omega_\gamma,
\label{spectr}
\end{eqnarray}
where the function $\lambda$ is defined by 
\begin{eqnarray}
\lambda (x,y,z)=\frac{(x-y-z)^2-4yz}{4x},
\end{eqnarray}
$M$ is the invariant mass of the $p{\bar p}$ system, $\Omega_p$ is the
proton angle in that system, while $\Omega_\gamma$ is the photon angle in
the $J/\Psi$ rest frame. After averaging over the spin states and 
integrating over the angles, the differential decay rate is
\begin{eqnarray}
\frac{d\Gamma}{dM}=\frac{(m_{J/\Psi}^2-M^2)\sqrt{M^2-4m_p^2}}
{2^7 \pi^3 m_{J/\Psi}^3}\,\, |A|^2 \ , 
\label{trans}
\end{eqnarray}
where $A$ is the total $J/\Psi{\to}\gamma p{\bar p}$ reaction 
amplitude. Note that $A$ is dimensionless.
The differential rate $d\Gamma{/}dM$, integrated over 
the range from $2m_p$ to $m_{J/\Psi}$, 
yields the partial $J/\Psi{\to}\gamma p{\bar p}$ decay width, 
$\Gamma{=}(3.3{\pm}0.9)10^{-2}$ keV~\cite{PDG}.
 
\begin{figure}[t]
\vspace*{-3mm}
\centerline{\hspace*{-1mm}\psfig{file=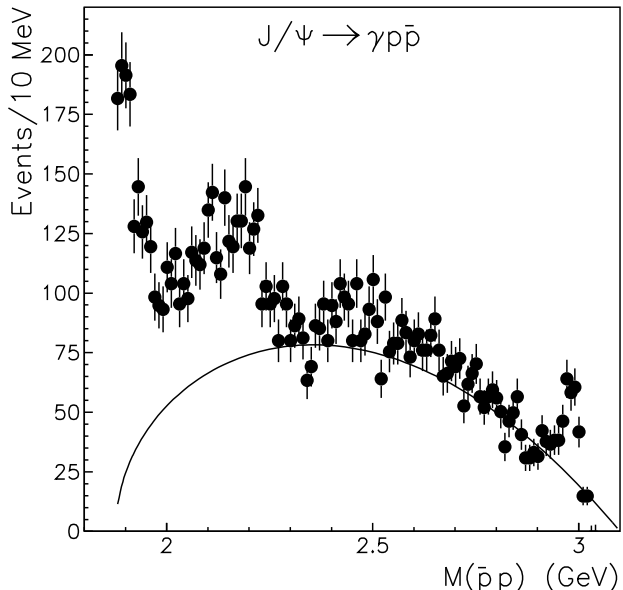,width=8.9cm,height=9cm}}
\vspace*{-7mm}
\caption{The $p{\bar p}$ mass spectrum from the decay $J/\Psi{\to}\gamma p{\bar p}$.
The circles show experimental results of the BES 
Collaboration~\cite{Bai}, while the solid line is the spectrum 
obtained from Eq.(\ref{trans}) by assuming a constant reaction amplitude 
$A$.}
\label{pbar5}
\end{figure}

The solid line in Fig. \ref{pbar5} is obtained with Eq.~(\ref{trans}) 
by using a constant $|A|^2$. This result corresponds to the
so-called phase space distribution. The actual value of $|A|^2$ was 
adjusted to the data at $M{\ge}2.8$ GeV, i.e. around the $\eta_c$ resonance 
region.  The circles in Fig. \ref{pbar5} show data 
for the $J/\Psi{\to}\gamma p{\bar p}$ decay published by the 
BES Collaboration~\cite{Bai}. Evidently, close to the $p{\bar p}$ threshold 
the data deviate substantially from the phase space distribution. 

Since close to the threshold the phase space factor introduces
a strong but trivial energy dependence of the mass distribution 
it is convenient 
to divide the experimental $p{\bar p}$ spectrum by the kinematical
factors that appear on the right side of Eq.(\ref{trans}) and extract the 
invariant amplitude $|A|^2$ from the data. 
Corresponding results for the BES data~\cite{Bai} are shown by
the squares in Fig.~\ref{pbar4}. It is obvious that the experimental
reaction amplitude exhibits a very strong mass dependence near
the $p{\bar p}$ threshold. 
In recent investigations \cite{Kerbikov2,Zou,Bugg} this feature has 
been attributed  to a strong final state interaction (FSI) between the 
outgoing protons and antiprotons. Indeed, an enhancement and/or a
strong energy dependence of near-threshold cross sections has been seen 
in many other reactions involving hadrons and is commonly
seen as to be caused by FSI effects, cf. 
\cite{Gasparian1,Sibirtsev3,Sibirtsev4,Nakayama,Baru,Hanhart,Hinterberger}.
Thus, it is plausible and it even has to be expected that FSI effects
play a role in the decay $J/\Psi{\to}\gamma p{\bar p}$ as well. 
The interesting issue is primarily whether the proton-antiproton interaction
is really strong enough to induce such a sizeable enhancement in the
invariant mass spectrum. 

A very simple and therefore also very popular treatment of 
FSI effects is due to Watson~\cite{Watson} and Migdal~\cite{Migdal}.
These authors suggested that the reaction amplitude 
for a production and/or decay reaction which is of short-ranged
nature can be factorized in terms of an elementary production 
amplitude $A_0$ and the $p\bar p$ scattering amplitude $T$ 
of the particles in the final state, 
\begin{eqnarray}
A_{prod} \approx N A_0 \cdot T,
\label{fsi}
\end{eqnarray}
where $A_0$ is  given by diagrams like those shown
in Fig. \ref{diag} and $N$ is a normalization factor.  

If the production mechanism is of short range then the 
amplitude $A_0$ depends only very weakly on
the energy and the near-threshold energy dependence of the 
reaction amplitude is driven primarily by the scattering 
amplitude, $T$, of the outgoing particles. 
This means that in the case of the reaction $J/\Psi{\to}\gamma p{\bar p}$ 
the near threshold mass dependence 
of the $p{\bar p}$ spectrum should  be dominated by the 
energy dependence of the $p{\bar p}$ scattering amplitude. 
In the analyses of the BES data by Kerbikov et al. \cite{Kerbikov2} and 
by Bugg \cite{Bugg} the above treatment of FSI effects was adopted.

However, the prescription of Watson--Migdal is only valid
for interactions that yield a rather large scattering length, 
like the $^1S_0$ $NN$ partial wave where the scattering length $a$ is 
in the order of 20 fm, and even then only for a relatively small energy
range, as was pointed out 
in several recent papers \cite{Hanhart2,Baru2,Gasparyan2}. 
Therefore, in the present case, where the scattering lengths are
in the order of 1 fm one should be cautious with the interpretation 
of results obtained from applying Eq.(\ref{fsi}).
Rather one should start from the more general expression for
the reaction amplitude, 
\begin{eqnarray}
A_{prod} = A_0  + A_0 \,G^{p\bar p} \,T,
\label{fsidwba}
\end{eqnarray}
which corresponds to a distorted wave born approximation. If one
assumes  that the production amplitude, $A_{prod}$, has only
a very weak energy- and momentum dependence it can be
factorized and one obtains
\begin{eqnarray}
A_{prod} \approx A_0 [1 + G^{p\bar p} T] = A_0 
\Psi^{(-)*}_{q_p}(0),
\label{fsidw}
\end{eqnarray}
where $\Psi^{(-)*}_{q_p}({\bf r})$ is a suitably normalized 
$N\bar N$ continuum wave function and $\Psi^{(-)*}_{q_p}(0)$
is nothing else but the inverse of the Jost function ${\cal J}$,
i.e. $\Psi^{(-)*}_{q_p}(0){=}{\cal J}^{-1}(-q_p)$ \cite{Goldberger}.
Eq.(\ref{fsidwba}) itself can be cast into the form \cite{Hanhart2}
\begin{eqnarray}
A_{prod} = A_0 [1  + (c-iq_p) T],
\label{fsidw2}
\end{eqnarray}
where $c$ is, in general, a complex number that represents the
principal-value integral over the half-off-shell extension of the  
production amplitude $A_0$ and the $p\bar p$ scattering amplitude
that appears on the very right hand side of Eq.(\ref{fsidwba}).
Obviously, Eq.(\ref{fsidw2}) is formally equivalent to the 
Watson-Migdal prescription of Eq.(\ref{fsi}) if $|c|$ is large. 
In case of the $NN$ $^1S_0$ partial wave $|c|$ is indeed large, as
has been demonstrated in Ref. \cite{Baru2}. 
%We should emphasize 
%though that $c$ is, of course, a model dependent quantity. Moreover, 
%$c$ is also not a constant but will depend on the energy (or the
%on-shell momentum $q_p$) too. However, for a short-ranged production
%operator with a very weak energy dependence it can be assumed
%that the energy dependence of $c$ is likewise weak. 

In considering the $S$-wave interaction 
in the final $p{\bar p}$ system one should account for the 
centrifugal barrier between the photon and the $p{\bar p}$ state 
given by the factor
\begin{eqnarray}
C_l = \left[\frac{m_{J/\Psi}^2-M^2}{2m_{J/\Psi}^2}\right]^{l_\gamma},
\end{eqnarray}
with $l_\gamma$=1 being the orbital momentum between the photon 
and the $p{\bar p}$ system. Around the threshold, $M{=}2m_p$, 
the centrifugal correction depends only weakly on the invariant mass 
of the $p{\bar p}$ system and basically does not modify the mass dependence 
of the FSI. Within the range 
$2m_p{\le}M{\le}2.1$~GeV the factor $C_1^2$ varies between 0.23 
and 0.27, which is not large enough to counterbalance the contribution 
from the strong $S$-wave FSI over the same invariant-mass range.

Note also that the Watson-Migdal formula (\ref{fsi}) has to be modified
when used for $p{\bar p}$ FSI effects in a $P$-wave. Since in the
production reaction only the final momentum is on-shell one has to
divide the on-shell $T$ matrix by the factor $q_p$ in order to 
impose the correct threshold behavior of the production amplitude. 

\section{Results}

\begin{figure}[t]
\vspace*{-3mm}
\centerline{\hspace*{-1mm}\psfig{file=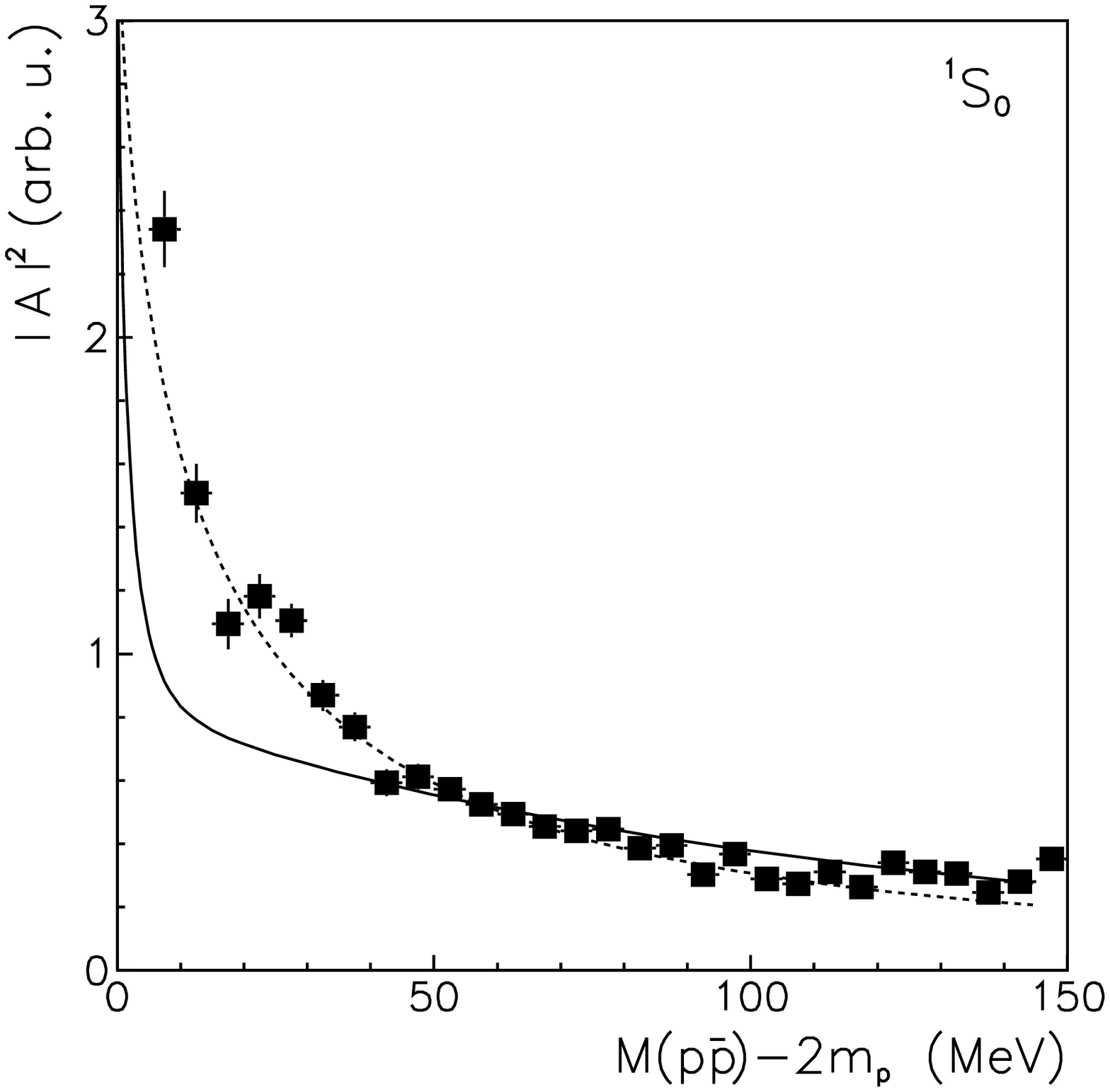,width=9.1cm,height=6.1cm}}
\vspace*{-6mm}
\centerline{\hspace*{-1mm}\psfig{file=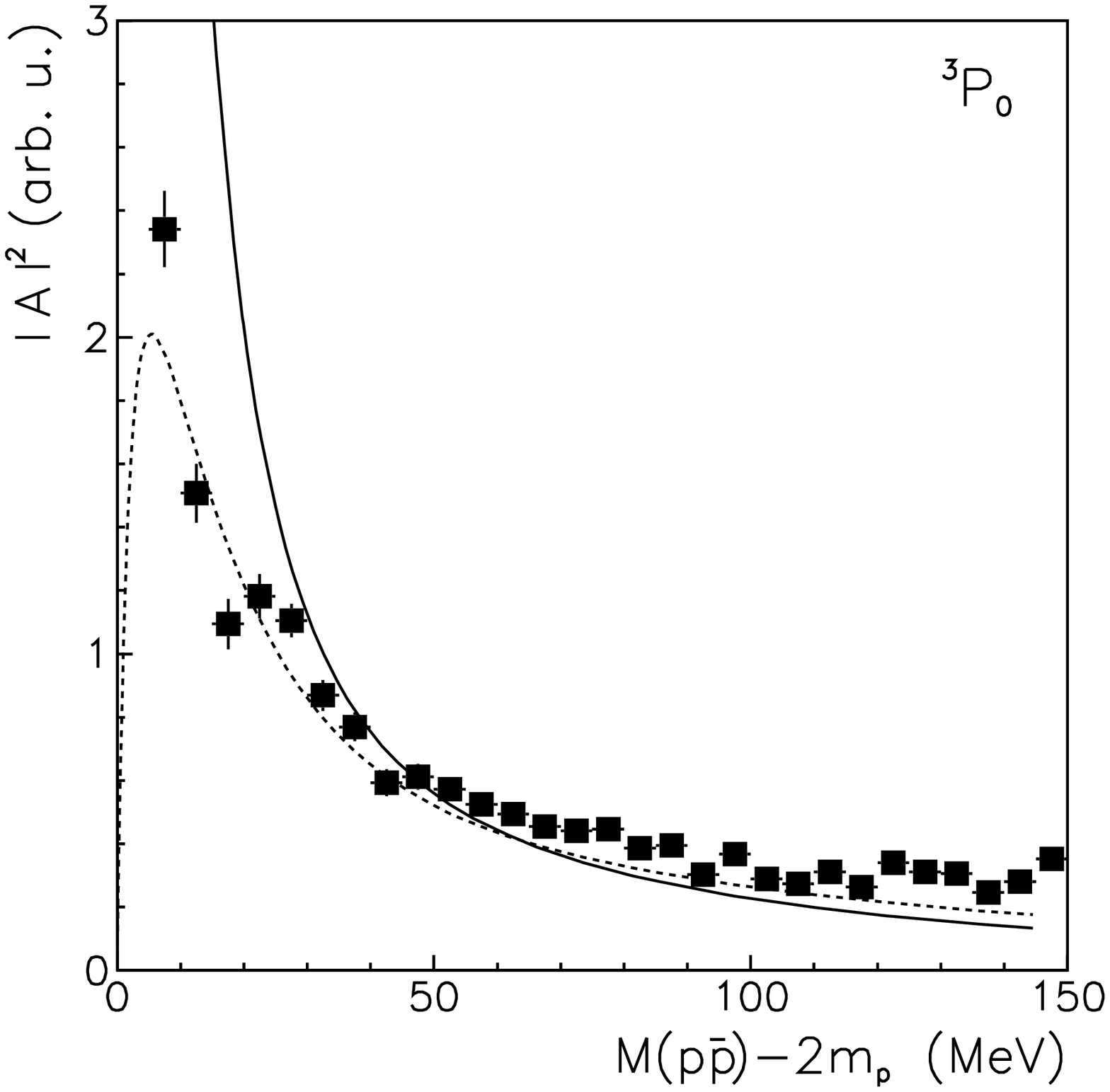,width=9.1cm,height=6.1cm}}
\vspace*{-4mm}
\caption{Invariant $J/\Psi{\to}\gamma p{\bar p}$ amplitude $|A|^2$ as a 
function of the $p{\bar p}$ mass. The squares represent the experimental 
values of $|A|^2$ extracted from the BES data\cite{Bai} via Eq.(\ref{trans}). 
The curves are the scattering amplitude squared ($|T|^2$) 
predicted by the $N\bar N$ model A(OBE) for the $^1S_0$ and $^3P_0$ partial 
waves and the $I{=}0$ (solid) and $I{=}1$ (dashed) channels,
respectively. Note that the latter results have been normalized to 
$|A|^2$ at $M(p\bar p){-}2mp{=}$50 MeV.}
\label{pbar4}
\end{figure}

Let us first discuss calculations based on the Watson-Migdal approach
given by Eq.~(\ref{fsi}).
The solid lines in Fig. \ref{pbar4} show the $p{\bar p}$ 
invariant scattering amplitudes squared for the $^1S_0$ and $^3P_0$ partial 
waves and the $I{=}0$ and $I{=}1$ channels. 
We consider  the isospin channels separately because, as mentioned above, 
the actual isospin
mixture in the final $p{\bar p}$ system depends on the reaction 
mechanism and is not known. Note that all squared $p\bar p$ 
scattering amplitudes
$|T|^2$ were normalized to the BES data at the invariant mass
$M(p{\bar p}){-}2m_p${=}50 MeV by multiplying them with a suitable 
constant.
The results indicate that the mass dependence of the BES data
can  indeed be described with FSI effects induced by 
the $^1S_0$ scattering 
amplitude in the $I{=}1$ isospin channel. The $I{=}0$ channel
leads to a stronger energy dependence which is not in agreement with the
BES data. We can also exclude dominant FSI effects from 
the $^3P_0$ partial waves. 
Here the different threshold behaviour due to the $P$-wave nature cannot
be brought in line with the data points very close to threshold. 
It should be clear, of course, that a suitable 
combination of several partial waves might as well reproduce the
experimental results on the $p{\bar p}$ invariant mass spectrum.

Note that we do not include the Coulomb interaction in our model calculation
and we also ignore the difference in the $p\bar p$ and $n\bar n$ thresholds. 
Judging from the results shown by Kerbikov et al. \cite{Kerbikov2} their
influence is noticeable only for excess energies below say 5 MeV. 
Accordingly we do not consider the lowest data point of the BES experiment,
which lies in the first (5 MeV) energy bin, in our discussion. 

\begin{figure}[t]
\vspace*{-4mm}
\centerline{\hspace*{-2mm}\psfig{file=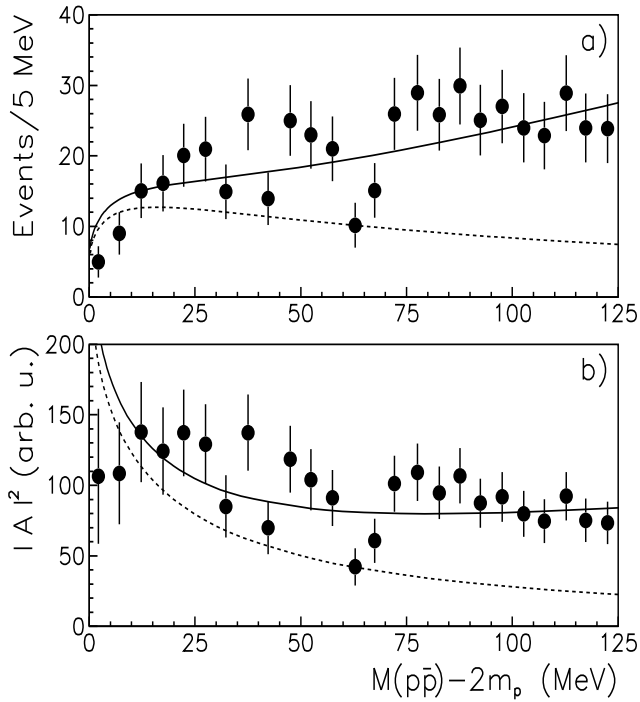,width=8.9cm,height=10.5cm}}
\vspace*{-6mm}
\caption{(a) The $p{\bar p}$ mass spectrum from the decay $J/\Psi{\to}\pi^0 
p{\bar p}$. The circles show experimental results of the BES 
Collaboration~\cite{Bai}.
(b) Invariant $J/\Psi{\to}\pi^0 p{\bar p}$ amplitude $|A|^2$ as a 
function of the $p{\bar p}$ mass. The circles represent the experimental 
values of $|A|^2$ extracted from the BES data via Eq.~(\ref{trans}). 
The curves are corresponding calculations using the $T$ matrix 
predicted by the $N\bar N$ model A(OBE) for the $^{33}S_1$ partial 
wave. The dashed line is the results from the Watson-Migdal prescription
Eq.~(\ref{fsi}), while the solid line is based on Eq.~(\ref{fsidw2})
with $c{=}-0.1$. Both curves are normalized arbitrarily.
}
\label{be8}
\end{figure}

Our results  support the conjecture of 
Kerbikov \cite{Kerbikov2} and Bugg \cite{Bugg} that the enhancement seen 
in the near-threshold
$p\bar p$ invariant mass spectrum of the decay $J/\Psi{\to}\gamma p{\bar p}$
could be primarily due to FSI effects in the $p\bar p$ channel. 
But we should also consider the 
$p\bar p$ invariant mass spectrum of the decay $J/\Psi{\to}\pi^0p{\bar p}$
which was presented by the BES collaboration in the same paper \cite{Bai}. 
Here the near-threshold mass distribution does not show any enhancement
as compared to the phase-space distribution -- though one would likewise
expect strong FSI effects in the $p\bar p$ channel. In this reaction 
isospin is preserved and the possible partial waves in the final
$p\bar p$ state, $^{33}S_1$ and $^{31}P_1$, differ from those 
available in the 
decay $J/\Psi{\to}\gamma p{\bar p}$. But the $^{33}S_1$ partial wave 
amplitude of the J\"ulich $N\bar N$ model, shown in the lower part of 
Fig. \ref{pbar6},
leads to FSI effects that are comparable to those of the $^{31}S_0$
state -- and in contradiction with the experimental $\pi^0p{\bar p}$ mass 
distribution - as  illustrated in  Fig.~\ref{be8}.
We should mention in this context that the issue of the
$\pi^0p{\bar p}$ mass distribution is not discussed in the work 
of Bugg, while Kerbikov argues that in the effective range parameterization
that he employs $|Re \ a_1|$ is much smaller than $|Re \ a_0|$ and 
therefore there should be much smaller FSI effects in the $I{=}1$ channel.
However, our experience is that $p\bar p$ amplitudes with a small $|Re \ a|$
often yield an even stronger energy dependence and therefore larger FSI 
effects - see the case of the $^{11}S_0$ partial wave in Fig.~\ref{pbar6}.

In our opinion these seemingly contradictory results can only be
reconciled if we assume that the treatment of FSI effects by means 
of Eq.(\ref{fsi}) is oversimplified and one should use
Eq.(\ref{fsidw2}) instead. Results based on the latter equation are 
shown by the solid lines in Fig.~\ref{be8} where we  set
$c{=}-0.1$. It might be possible that a fine
tuning of $c$ could allow one to better reproduce the BES data for 
the $\pi^0 p\bar p$ channels, though in view of the rather large 
statistical variations in the data we refrain from doing so. 
In any case it is obvious that the more refined treatment of the FSI effects 
based on Eq. (\ref{fsidw2}) leads, in general, to a weaker energy dependence 
of the $p\bar p$ mass spectrum. 

\section{Discussion}

It is important to note that the near-threshold contribution to the 
$J/\Psi{\to}\gamma p{\bar p}$ decay rate is 
relatively large. The contribution to $J/\Psi{\to}\gamma p{\bar p}$ 
up to invariant $p\bar p$ masses of $M(p\bar p){\approx}$ 2 GeV 
is roughly 5 times larger than the rate for
the reaction $J/\Psi{\to}\gamma\eta_c$ followed by the $\eta_c{\to}p{\bar p}$ 
decay.  Taking into account that the latter rate was recently 
published~\cite{Bai1} as BR=$1.9{\cdot}10^{-5}$ we can estimate that
\begin{eqnarray}
BR(J/\Psi{\to}\gamma ^1\tilde S_0)\times BR(^1\tilde S_0{\to}p{\bar p})\simeq
9.5\cdot 10^{-5},
\end{eqnarray}
where $^1\tilde S_0$ indicates the near-threshold contribution of this partial 
wave -- 
which is indeed a large fraction of the total $J/\Psi{\to}\gamma p{\bar p}$
branching rate that amounts to $(3.8{\pm}1.0){\cdot}10^{-4}$.
 
\begin{figure}[b]
\vspace*{-6mm}
\centerline{\hspace*{1mm}\psfig{file=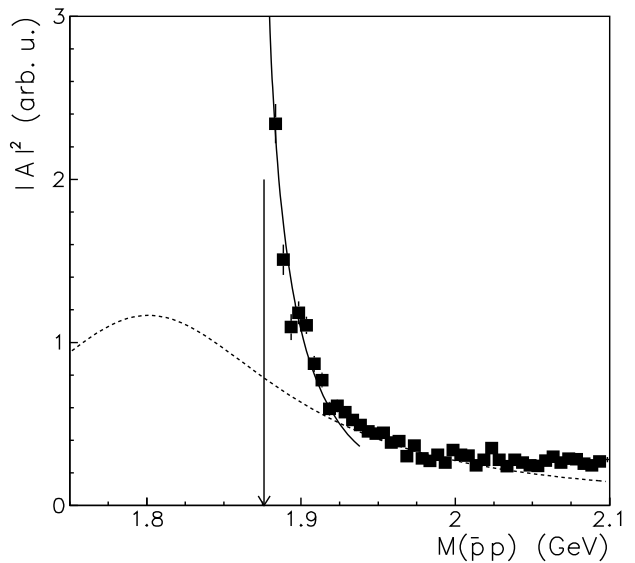,width=8.7cm,height=8.4cm}}
\vspace*{-5mm}
\caption{Invariant $J/\Psi{\to}\gamma p{\bar p}$ amplitude $|A|^2$ as 
a function of the $p{\bar p}$ mass.  The squares represent the experimental 
values of $|A|^2$ extracted from the BES data\cite{Bai} via Eq.(\ref{trans}). 
The dashed line
shows the $\pi$(1800) resonance amplitude squared, while the solid line
indicates the square of the product of resonance and $p{\bar p}$
$^1S_0$ scattering amplitude given by Eq.(\ref{fsi}). The arrow shows the
$p{\bar p}$ threshold.}
\label{pbar3}
\end{figure}

This suggests that in this energy region the reaction 
$J/\Psi{\to}\gamma p{\bar p}$ 
could be indeed dominated by a resonance. Such a resonance should
lie below the $p\bar p$ threshold but, unlike the resonance state which
emerged from the Breit-Wigner fit of the BES collaboration \cite{Bai}, 
its mass could be significantly below the $p{\bar p}$ threshold and it 
could have a large width, like the $\pi(1800)$ or $\eta(1760)$ resonances
that are listed in the Particle Physics Booklet \cite{PDG}. 
We demonstrate the scenario in Fig.~\ref{pbar3}.  
For illustration we consider the $\pi$(1800) meson 
with mass and width 1801 and 210 MeV, respectively. The corresponding
resonance amplitude squared is shown by the dashed line in Fig.~\ref{pbar3}. 
Its energy dependence is, of course, too weak and does not agree with
the near-threshold $p\bar p$ mass spectrum found by the BES collaboration. 
However, when the $\pi$(1800) resonance decays into a proton and antiproton, 
a final state interaction should occur. The solid line in Fig.~\ref{pbar3} 
shows the total reaction amplitude squared, taken now as the product of the 
resonance amplitude and the $p{\bar p}$ scattering amplitude in the $^{31}S_0$ 
partial wave as is given by Eq.(\ref{fsi}). The corresponding result
is roughly in line with the BES data.

Though this line of reasoning is certainly more a 
plausibility argument, rather than
a solid calculation, we believe that it makes clear that one has to
be rather cautious when trying to extract resonance properties by fitting 
the near threshold $p{\bar p}$ mass spectrum with a resonance amplitude, 
because any final state interaction will necessarily and substantially 
distort the production amplitude. 

A different and alternative scenario consists in assuming 
the formation of many multimesonic intermediate
states in the $J/\Psi$ decay that couple strongly to the $p{\bar p}$ system
where again a strong FSI occurs. The branching rates of the $J/\Psi$ 
radiative decays into the $\pi^+\pi^-2\pi^0$, $\rho\rho$, $2\pi^+2\pi^-$
and $\omega\omega$ channels are larger than $10^{-3}$ and the transition of 
those mesonic states into the $p{\bar p}$ final state could be sufficiently
strong~\cite{Hippchen,Mull}. Thus, such a scenario cannot be excluded
by the available data.

Finally, let us come back to the other proposed explanation for the
enhancement in the $p{\bar p}$ mass spectrum, namely near-threshold
$N\bar N$ bound states. Clearly a description of the experimental
mass spectrum in terms of $p{\bar p}$ FSI effects does not contradict
the existence of such states. 
Indeed, in case of the $NN$ interaction the strong FSI effects are
interconnected with the existence of a near-threshold bound state
(deuteron) or anti bound-state in the corresponding $^3S_1$ and 
$^1S_0$ partial waves. However, the $N\bar N$ model that we used
in the present study does not lead to any near-threshold bound
states. In fact, we found only two bound states for the model A(OBE) 
within 100 MeV from the threshold, namely at 
$E$ = --104--$i$413 MeV in the
$^{11}S_0$ partial wave and at $E$ = --24.2--$i$107 MeV in the 
$^{13}P_0$ partial wave. Obviously both bound states lie rather
far away from the real axis and therefore should have practically
no influence on the $p{\bar p}$ scattering amplitude
near threshold.

\section{Conclusion}
We have investigated suggested explanations for the near-threshold 
enhancement in the $p{\bar p}$ invariant mass spectrum of the 
$J/\Psi{\to}\gamma p{\bar p}$ decay, reported recently 
by the BES Collaboration. In particular, we showed that the near-threshold 
enhancement in the $p{\bar p}$ mass spectrum can, in principle, 
be understood in terms of a final state interaction in the outgoing 
proton-antiproton system. Within our model calculation it can be described 
with the scattering amplitude in the $^{31}S_0$ $p{\bar p}$ partial wave but 
disagrees with the energy dependence that follows from the 
$^{11}S_0$ and both ($I=0,1$) $^3P_0$ amplitudes.

We showed that the scattering length approximation cannot 
be used for a more quantitative evaluation of the energy dependence 
of BES data.
In general, it does not reproduce the 
energy dependence of the scattering amplitude reliably enough within 
the $p{\bar p}$ mass range required.
At the same time, one has to concede that the lack of knowledge of the
reaction mechanism for $J/\Psi{\to}\gamma p{\bar p}$ and the insufficient
information on the $p{\bar p}$ interaction near threshold precludes any
more quantitative conclusions, even when microscopic models are used
- as in  the case of the present study. 

We argued also that the use of the simple Watson--Migdal prescription for 
treating FSI effects has to be considered with caution for the present case 
in view of the small $p\bar p$ scattering lengths, whose real parts are 
typically in the order of only one fermi. 
Our suspicion is nourished, in particular, by the observation that the 
experimental $p{\bar p}$ mass spectrum of the comparable decay
$J/\Psi{\to}\pi^0 p{\bar p}$ does not show any obvious sign of a
$p{\bar p}$ FSI, while applicaton of the Watson--Migdal prescription 
would yield a similar, strong, near-threshold enhancement as for the
$\gamma p{\bar p}$ channel for any of the $N\bar N$ models we 
utilized. 
We demonstrated that a consistent qualitative description of the
$p{\bar p}$ invariant mass spectra from the decay reactions
$J/\Psi{\to}\gamma p{\bar p}$ and $J/\Psi{\to}\pi^0 p{\bar p}$
can be achieved, however, within a more refined treatment of FSI 
effects as it follows from a DWBA approach. 

Though our study shows that the enhancement seen in the decay
$J/\Psi{\to}\gamma p{\bar p}$ could indeed be a result of the FSI in the
$p{\bar p}$ system one has to admit that due to the uncertainties 
mentioned above and the controversal situation in the $\pi^0 p{\bar p}$ 
channel explanations other than final state effects cannot be ruled out at
the present stage.   
Since the available data on the reaction $J/\Psi{\to}\pi^0 p{\bar p}$
are afflicted by large error bars it would be desirable to obtain improved 
experimental information here that allows one to quantify the extent of
$p\bar p$ FSI effects in this reaction. 
As already mentioned in the
introduction, other reactions involving the $p\bar p$ system in the 
final state, such as $B^+{\to}K^+p{\bar p}$ decay~\cite{Abe1}
or ${\bar B}^0{\to}D^0p{\bar p}$~\cite{Abe2} do show some indications
for $p\bar p$ FSI effects. But here too the quality of the available
data is too poor to permit any reliable conclusions. 
 
In this context let us emphasize that $p{\bar p}$ mass spectra for
the reactions $J/\Psi{\to}\omega p{\bar p}$ or $J/\Psi{\to}\eta p{\bar p}$
would be rather interesting for clarifying the role of $p\bar p$ FSI effects
in the decay of the $J/\Psi$ meson. Both reactions restrict the isospin 
in the $p{\bar p}$ system to be zero so that one could explore the 
FSI for specific $p\bar p$ partial waves, namely 
$^{11}S_0$ and $^{13}S_1$. 

\acknowledgments{
We would like to thank D. Diakonov, A. Kudryavtsev,
G. Miller, F. Myhrer and W. Schweiger for useful discussions.
This work was partially supported by the Department of
Energy under contract DE-AC05-84ER40150 under which SURA operates
Jefferson Lab, and  by 
grant No. 447AUS113/14/0 of the Deutsche Forschungsgemeinschaft 
and the Australian Research Council.
}

\end{document}